# A Transfer Learning Based Active Learning Framework for Brain Tumor Classification

**Ruqian Hao[1,2,5], Khashayar Namdar[3,5], Lin Liu[1], Farzad Khalvati[2,3,4,5]**

[1]School of Optoelectronic Science and Engineering, University of Electronic Science and Technology of China, China

[2]Institute of Medical Science, University of Toronto, Toronto, ON, Canada

[3]Department of Medical Imaging, University of Toronto, Toronto, ON, Canada

[4]Department of Mechanical and Industrial Engineering, University of Toronto, Toronto, ON, Canada

[5]The Hospital for Sick Children (SickKids), Toronto, ON, Canada

**Abstract**

Brain tumor is one of the leading causes of cancer-related death globally among children and adults. Precise classification of brain tumor grade (low-grade and high-grade glioma) at early stage plays a key role in successful prognosis and treatment planning. With recent advances in deep learning, Artificial Intelligence-enabled brain tumor grading systems can assist radiologists in the interpretation of medical images within seconds. The performance of deep learning techniques is, however, highly depended on the size of the annotated dataset. It is extremely challenging to label a large quantity of medical images given the complexity and volume of medical data. In this work, we propose a novel transfer learning based active learning framework to reduce the annotation cost while maintaining stability and robustness of the model performance for brain tumor classification. We employed a 2D slice-based approach to train and finetune our model on the Magnetic Resonance Imaging (MRI) training dataset of 203 patients and a validation dataset of 66 patients which was used as the baseline. With our proposed method, the model achieved Area Under Receiver Operating Characteristic (ROC) Curve (AUC) of 82.89% on a separate test dataset of 66 patients, which was 2.92% higher than the baseline AUC while saving at least 40% of labeling cost. In order to further examine the robustness of our method, we created a balanced dataset, which underwent the same procedure. The model achieved AUC of 82% compared with AUC of 78.48% for the baseline, which reassures the robustness and stability of our proposed transfer learning augmented with active learning framework while significantly reducing the size of training data.

## 1 Introduction

Brain tumor is one of the leading causes of cancer-related death globally among children and adults (Siegel, Miller, and Jemal 2019). According to the World Health Organization (WHO) classification 2016 (Louis et al. 2016), brain tumors are divided into different grades (grade I, II, III, or IV) based on histology and molecular characteristics. The higher the grade of the tumor is, the more malignant it becomes. Patients with Low Grade Glioma (LGG, grade I/II) usually have better survival than those diagnosed with High Grade Glioma (HGG, grade III/IV), which is incurable and universally fatal. LGGs have high possibility of eventually progressing to HGG if it is not diagnosed and the treatment is delayed (Claus et al. 2016).

Precise classification of brain tumor grade at the early stage plays a key role in successful prognosis (Delattre et al. 2014). Magnetic Resonance Imaging (MRI) is the favored imaging technique in gliomas diagnostics due to good contrast enhancement and non-invasive features (Essig et al. 2012). The conventional method for tumor detection is by radiologists who observe and diagnose tumors which is extremely laborious and time-consuming. Recent advances in Artificial Intelligence (AI) and deep learning have made great strides in Computer-Aided Medical Diagnosis (CAMD), which can assist doctors in the interpretation of medical images within seconds (Hosny et al. 2018).

The performance of deep learning technique is highly dependent on the quality and size of the dataset. Deep learning techniques require large number of images with high-quality annotations. However, labelling large quantities of medical images is quite challenging as annotation can be expensive in terms of both time and expertise (Razzak, Naz, and Zaib 2018). For example, even an expert neuroradiologist with more than ten years of experience may need hours to correctly label the brain tumor image of one patient (Fiez, Damasio, and Grabowski 2000). Insufficient amount of imaging data and scarcity of human expert annotations for images are the two major barriers in success of deep learning for medical imaging (Razzak, Naz, and Zaib 2018).

To address and resolve the abovementioned challenges, numerous efforts have been made. For instance, transfer learning is a promising strategy in case of limited domain training samples. It finetunes a network which is already pretrained on a large labeled dataset, typically from another domain. By transferring learned knowledge to the target small dataset, the speed of network convergence becomes faster while maintaining low computational complexity level at the training stage (Tajbakhsh et al. 2016).

Active learning algorithms have also been investigated to train a competitive classifier with minimal annotation cost. The underlying idea behind active learning is that different training examples have different effects on the performance of the current model. Instead of labeling the complete dataset, an active learning method selects a subset of informative samples to annotate and then train the classification model without compromising its performance. There are two important metrics to describe the informativeness of an unlabeled sample: uncertainty, which is the inverse of the confidence of predicted results by the model; and representativeness, which measures the degree of similarity in distribution and structure between selected samples and target dataset (Du et al. 2017). Based on different query schemes of informative unlabeled samples, conventional active learning algorithms can be listed as follows: uncertainty sampling, query-by-committee, expected model change, expected error reduction, variance reduction and density-weighted methods (Settles 2011).

In this work, we propose an active learning method which integrates traditional uncertainty sampling technique and query-by-committee method, and transfer learning to reduce the amount of required training samples while maintaining stability and robustness of Convolutional Neural Network (CNN) performance for brain tumor classification.

## 2  Materials and Methods

### 2.1  Related Work

#### 2.1.1 Brain Tumor classification using deep learning

(Mohsen et al. 2018) used a Deep Neural Network (DNN) combined with the Discrete Wavelet Transform (DWT) to classify brain MRIs collected from Harvard Medical School website (Summers 2003) into four types which are tumor-free, glioblastoma, sarcoma, and metastasis, and the classification accuracy was 96.97%. (Pereira et al. 2018) proposed a novel CNN with deeper architectures and small Kernels for automatic LGG and HGG brain tumor grading prediction on both whole brain and only tumor region MRI images, and the accuracies were 89.5% and 92.98, respectively. The datasets they used are BRATS 2013 and BRATS 2015. (Suganthe et al. 2020) employed a Recurrent Neural Network (RNN) architecture for detection of tumors on a 600 MRI brain images dataset and achieved an accuracy of 90%. On a brain tumor dataset consisting of 3,064 MRI images from 233 patients, there has been multiple experiments (Badža and Barjaktarović 2020)(Sunanda Das, O.F.M. Riaz Rahman Aranya 2019)(Afshar, Mohammadi, and Plataniotis 2018). Each patient in the dataset has one of three types of brain tumor (glioma, meningioma and pituitary). (Badža and Barjaktarović 2020) presented a new CNN architecture for three types of brain tumor classification and the best accuracy was 96.56%. (Sunanda Das, O.F.M. Riaz Rahman Aranya 2019) also explored a CNN model for classification of three types of brain tumor MRI images and an accuracy of 94.39% was achieved. (Afshar, Mohammadi, and Plataniotis 2018) proposed a modified CapsNet architecture (Ballal and Zelina 2004) combined with tumor boundaries information for brain tumor classification and achieved 90.89% accuracy.

### 2.1.2 Transfer learning and active learning for medical imaging

(Yang et al. 2018) compared the classification performance of finetuned pretrained CNNs and CNNs trained from scratch on a private glioma MRI dataset containing 113 LGG and HGG patients. The experiments showed that transfer learning and finetuning improved performance for classifying HGG and LGG. They achieved their best test accuracy of 90%, using GoogLeNet. (Banerjee et al. 2019) proposed three CNN models (PatchNet, SliceNet, and VolumeNet), trained from scratch and compared with the two pretrained ConvNets (VGGNet (Simonyan and Zisserman 2015) and ResNet (Li et al. 2019)) finetuned on the BRATS 2017 dataset for HGG and LGG classification problem. Results demonstrate that the proposed VolumeNet achieved best testing accuracy of 95%. (Swati et al. 2019) used a block-wise finetuning algorithm based on transfer learning to finetune pretrained CNN on a MRI brain tumor dataset and obtained average accuracy of 94.82% under five-fold cross validation. (Rehman et al. 2019) employed three pretrained CNNs (AlexNet (Krizhevsky 2007), GoogLeNet (Zeng et al. 2016), and VGGNet (Simonyan and Zisserman 2015)) to classify brain tumors MRI images with two different transfer learning techniques (finetune and freeze), and the finetune VGG16 architecture showed the highest accuracy of 98.69%.

(Smailagic et al. 2018) sampled the instances which had the longest distance from other training samples in a learned feature space. The proposed strategy reduced the annotated examples by 32% and 40% respectively, compared to the conventional uncertainty and random sampling methods on the task of Diabetic Retinopathy detection. (Dai et al. 2020) proposed a gradient-guided suggestive annotation framework which computes gradient of training loss and then selects informative examples which have the shortest Euclidean distance to the gradient-integrated samples projected onto the data manifold learned by a variational autoencoder (VAE). Through employing this framework, they selected 19% of the MRI images from BRATS 2019 dataset to train a CNN for brain tumor segmentation task and achieved competitive results (a Dice score of 0.853) compared with when the whole labeled dataset was used. (Zhou et al. 2017) augmented each sample by data augmentation technique, and then computed entropy and relative entropy for original and augmented samples. Next, they continuously selected the most uncertain samples to label and added them to the training dataset to finetune AlexNet at each iteration. They managed to cut the needed annotated training data by half in three different biomedical imaging applications. (Li et al. 2019) proposed an active learning strategy for breast cancer classification on pathological image dataset. Instead of selecting the most informative samples, the algorithm removed 4,440 misleading samples from the training dataset which contained 68,640 samples. They obtained patch-level average classification accuracy of 97.63%, compared to 85.69% which was resulted by training on the whole dataset.

## 2.2 Methods

Transfer learning is a widely used approach in which a network is trained on a large labeled (source) dataset and the resulting pretrained network is finetuned on the target small dataset, transferring the learned knowledge from the source to target dataset. Active learning, on the other hand, is a promising strategy which has been investigated to train a competitive classifier with minimal annotation cost. In this work, transfer learning and active learning are the components of our proposed uncertainty sampling method for achieving stable test results using a smaller subset of training cohort. We chose the MICCAI BRATS 2019 dataset (Menze et al. 2015)(Bakas et al. 2017)(Bakas et al. 2018) as the target dataset which is a new, well annotated, well preprocessed, and skull striped dataset with interpolation and registration.

### 2.2.1 Dataset

All the experiments in this work were performed on the BRATS 2019 dataset (Menze et al. 2015) which consists of 335 patients diagnosed with brain tumors (259 patients with HGG, 76 patients with LGG). Each patient MRI scan set has four MRI sequences, which are T1-weighted, post-contrast enhanced T1-weighted (T1C), T2-weighted (T2), and T2 Fluid-Attenuated Inversion Recovery (FLAIR) volumes. The dataset was preprocessed with skull-striping, interpolation to a uniform isotropic resolution of 1 mm3 and registered to SRI24 space with a dimension of $240 \times 240 \times 155$. The annotations of the dataset include four labels: background, Gadolinium-enhancing tumor, the peritumoral edema, and the necrotic and non-enhancing tumor core. The area identified by the last three of the four labels represents the complete tumor region.

To implement the proposed method in this work, we randomly extracted 20 slices with tumor region from each patient MRI scan in axial plane, and kept T1, T1C and T2 channels for each slice. The pretrained AlexNet requires three channel input, and we chose T1, T1C and T2 channels from total four channels based on the results of the initial experiments. The obtained 6,700 2D 3-channel slices dataset was further split into training set (203 patients), validation set (66 patients), and test set (66 patients). All the three cohorts have the same ratio of HGG patient number and LGG patient number as the full dataset. Every slice with LGG tumor was annotated as label 0, and HGG tumor slices were labeled as 1. The images were resized from 240×240 pixels to 224×224 pixels in order to fit the pretrained CNN.

### 2.2.2 Transfer Learning

Training a CNN from scratch (with random initialization) requires massive amount of annotated training samples and relatively more time and computational resources than employing a CNN pretrained on a very large dataset. In general, there are two main scenarios of transfer learning: finetuning and freezing. In finetuning, instead of random initialization, weights and biases of a pretrained CNN are adopted and then a conventional training process on the target dataset is performed. In the freezing scenario, we consider the pretrained CNN layers as a fixed feature extractor. In this context we freeze the weights and biases of our desired convolutional layers, and let the fully-connected layers be finetuned over the target dataset. The frozen layers do not have to be limited to the convolutional layers. Frozen layers can be chosen to be any subset of convolutional or fully-connected layers, however, common practice is to freeze the shallower convolutional layers. In our research, the CNNs are pretrained on ImageNet Large-Scale Visual Recognition Challenge (ILSVRC) dataset (Russakovsky et al. 2015) which includes natural images. Due to the large difference between our target medical image domain and the ImageNet dataset, we chose the finetuning to be our strategy of transfer learning.

Based on the purpose of reducing the annotation cost, we opted the pretrained AlexNet and finetuned it on the BRATS 19 dataset. AlexNet is composed of five convolutional layers, three max-pooling layers, and three fully connected layers. The detailed architecture used in this work is shown in Table 1. AlexNet depth is capable for brain tumor classification and it is considerably shallower than other benchmark CNNs (e.g., ResNet (He et al. 2016), VGG (Simonyan and Zisserman 2015)), which leads to faster convergence and less required computational resources.

Table 1. The detailed architecture of AlexNet

| Layer | Kernel size | Stride | Padding | Output size |
| --- | --- | --- | --- | --- |
| Conv1 | 11×11 | 4 | 2 | 64×55×55 |
| Maxpool1 | 3×3 | 2 | 0 | 64×27×27 |
| Conv2 | 5×5 | 2 | 2 | 192×27×27 |
| Maxpool2 | 3×3 | 2 | 0 | 192×14×14 |
| Conv3 | 3×3 | 1 | 1 | 384×13×13 |
| Conv4 | 3×3 | 1 | 1 | 256×13×13 |
| Conv5 | 3×3 | 1 | 1 | 256×13×13 |
| Maxpool3 | 3×3 | 2 | 0 | 256×6×6 |
| FC1 | | | | 4096×1 |
| FC2 | | | | 4096×1 |
| FC3 | | | | 2×1 |

**2.2.3 Uncertainty Score Calculation**

We use entropy and relative entropy as measures to estimate the informativeness of each training example. Given a discrete random variable X, with possible outcomes $x_1, x_2, \ldots, x_n$, which occur with probabilities $P(x_1), P(x_2), \ldots, P(x_n)$, the entropy formula of X is given by Equation (1).

$$H(X) = -\sum_{i=1}^{n} P(x_i) \log P(x_i) \tag{1}$$

Another useful measure for estimating the amount of mutual information between two possibility distributions on a random variable is relative entropy, also known as the symmetric Kullback-Leibler (KL) divergence. Formally, given two probability distributions $P(x)$ and $Q(x)$ over a discrete random variable X which has n possible outcomes, the relative entropy given by $D(p||q)$ is given by Equation (2).

$$D(p||q) = \sum_{i=1}^{n} P(x_i) \log \frac{P(x_i)}{Q(x_i)} \tag{2}$$

In this scenario, the probability distributions are the outputs of the pretrained CNNs.

### 2.2.4 Workflow

In this work, we present a novel transfer learning based active learning framework to reduce the annotation cost while maintaining stability and robustness of CNN performance for brain tumor classification. Our active learning workflow is described in Figure 1.

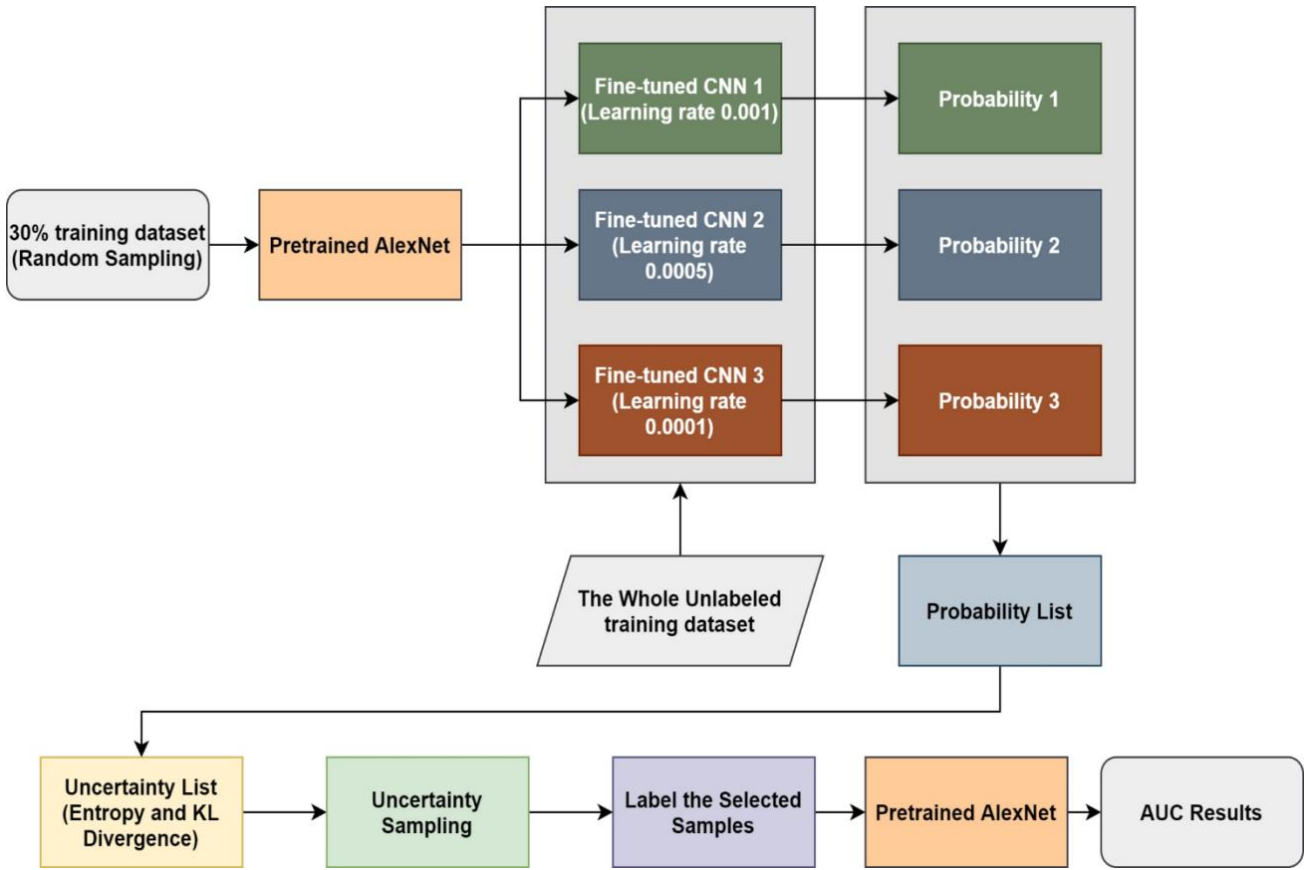

Figure 1. Workflow of Proposed Transfer Learning Based Active Learning Framework

We assume the training dataset consists of labeled and unlabeled subsets. The goal is to find the best informative samples in the entire training set, which may or may not overlap with the labeled training subset. The workflow is divided into four steps: 1) for the labeled training subset, we randomly selected 30% training samples and assumed the remaining 70% samples were unlabeled. We then used the 30% labeled training subset to finetune the pretrained AlexNet, and the learning rate α was set to different values (i.e., 0.001, 0.0005, and 0.0001). By performing this step, we obtained three finetuned CNNs. 2) we used these finetuned CNNs to compute the classification probabilities of each sample in the entire training dataset. In this step, the CNNs only perform forward propagation to calculate outputs, therefore no labels are required. 3) once each training sample produced three predicted possibilities in step 2, we computed the individual entropy (Equation 1) and pairwise KL divergence (Equation 2). The uncertainty score is the sum of the entropy and KL divergence of each sample. Through this approach, an uncertainty score list of the entire training dataset was obtained. 4) we sorted the uncertainty score list in descending order and we sampled 30% of the training cohort, which consisted of the best informative samples. This selected subset required labeling and was consequently used to finetune a pretrained AlexNet.

If there was no overlap between the original labeled training subset (30%) and the discovered best informative subset (30%), then the maximum training size needed is 30+30=60% (40% reduction in training size) of the entire training cohort. If the discovered best informative samples happen to be exactly the same as the original labeled training subset (30%), then the maximum training size needed is only 30% (70% reduction in training size) of the entire training cohort. In other words, between 40% and 70% of annotation cost (average of 55%) can be saved by our proposed transfer learning based active learning framework.

## 3 Results

All the experiments were conducted on a NVIDIA GeForce RTX 2070 platform, using Python 3.8 and PyTorch 1.5.1. In order to prove the stability and reproducibility, all the AUC results below are averages of 10 runs of a single experiment and presented as mean along with the 95% Confidence Interval (CI).

In section 3.1, we will show transfer learning is an effective approach and improves our baseline models. In section 3.2, we will demonstrate the top 10% certain and uncertain examples are not informative and thus, omitting them helps the models to better generalize. In section 3.3, we will experimentally show our uncertainty sampling approach improves the baseline with sample size fixed at 30%. Finally, in section 3.4, we will demonstrate: a) regardless if the dataset is balanced or imbalanced, our sampling method is effective b) the fact that our sampling approach improves the baseline is not arbitrary or as a result of filtering noisy examples through chance. It in fact always outperforms random sampling c) Although 30% is the optimum sample size, our sampling method works at other sample sizes as well.

### 3.1 Results of using Transfer Learning

Training AlexNet from scratch requires massive data with high-quality annotation. Employing transfer learning technique improves performance of the model when sufficient data is not available. The baseline AUC was computed by finetuning the pretrained AlexNet on the entire training dataset. The maximum number of epochs was 30, learning rate was set to 0.001, batch size was set to 16, momentum in Stochastic Gradient Descent (SGD) optimizer was 0.8 and L2 regularization penalty was set to 0.0001 based on a grid search strategy. We also explored training AlexNet from scratch, with the same hyperparameter settings except that epoch number was increased to 80 because it needed more iterations to converge.

Table 2 lists AUC results with and without transfer learning strategy on both validation dataset and test dataset. As it can be seen, the validation AUC and test AUC improved by 1.51% and 7.98% respectively when employing transfer learning method.

Table 2. AUC results of AlexNet trained from scratch and finetuned from pretrained model

| AUC (95%CI) | Pretrained AlexNet | AlexNet trained from scratch |
| --- | --- | --- |
| Validation dataset | 87.46% (87.11, 87.81) | 86.14% (85.60, 86.68) |
| Test dataset | 79.91% (78.95, 80.87) | 71.93% (70.76, 73.10) |

### 3.2 AUC Results of Selecting Different Range of Uncertainty Distribution

As described in section 2.2.4, we finetuned the pretrained AlexNet on 30% of the training dataset, which was labeled, and obtained three finetuned CNNs with learning rate α set to 0.001, 0.0005, and 0.0001, respectively. The uncertainty score list of the entire training samples was computed based on the output of these CNNs. Figure 2 visualizes uncertainty distribution of the training dataset, where uncertainty score list is unsorted in Figure 2(a), and uncertainty scores are ranked in descending order in Figure 2(b).

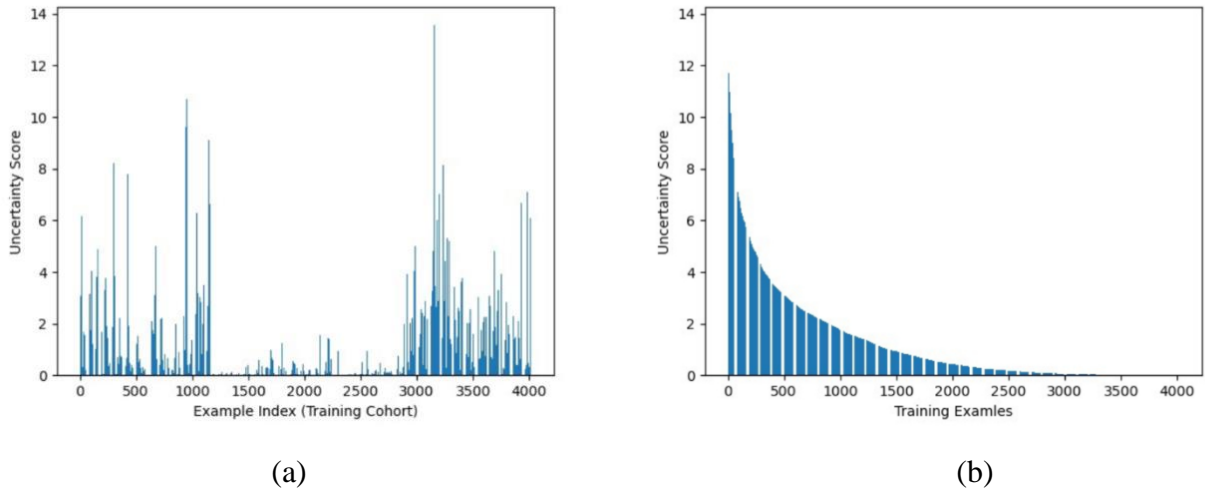

(a)  (b)

Figure 2. Visualization of uncertainty distribution of training dataset. (**a**) unsorted; (**b**)sorted

While keeping the number of samples constant (i.e., 30% of the training dataset), we finetuned the pretrained AlexNet on different ranges of uncertainty distribution. This was done to assess the effect of sampling from diverse uncertainty ranges on the performance of the CNN. In all experiments, we stopped our training or finetuning procedure at the highest validation AUC.

As reflected in Figure 3, AUC results for validation and test sets were calculated on samples from different uncertainty ranges according to the sorted uncertainty list. As it can be seen, the biggest jumps of validation AUC occur when the first and last 10% of the sorted list (the top 10% certain and uncertain examples) are excluded. As shown in Figure 3, using the top 30% certain examples or the top 30% uncertain examples results in a decrease of AUC results for validation (and test) cohort. Thus, we removed the top 10% (highest uncertainty scores) and the bottom 10% (lowest uncertainty scores) samples to eliminate outliers with least training values. As it can be seen in Figure 3, uncertainty range of 10%-40% improves AUC results by 12.51% compared to the range of 0-30%. Similarly, uncertainty range of 60%-90% elevates AUC by 7.72% in comparison to the range of 70%-100%.

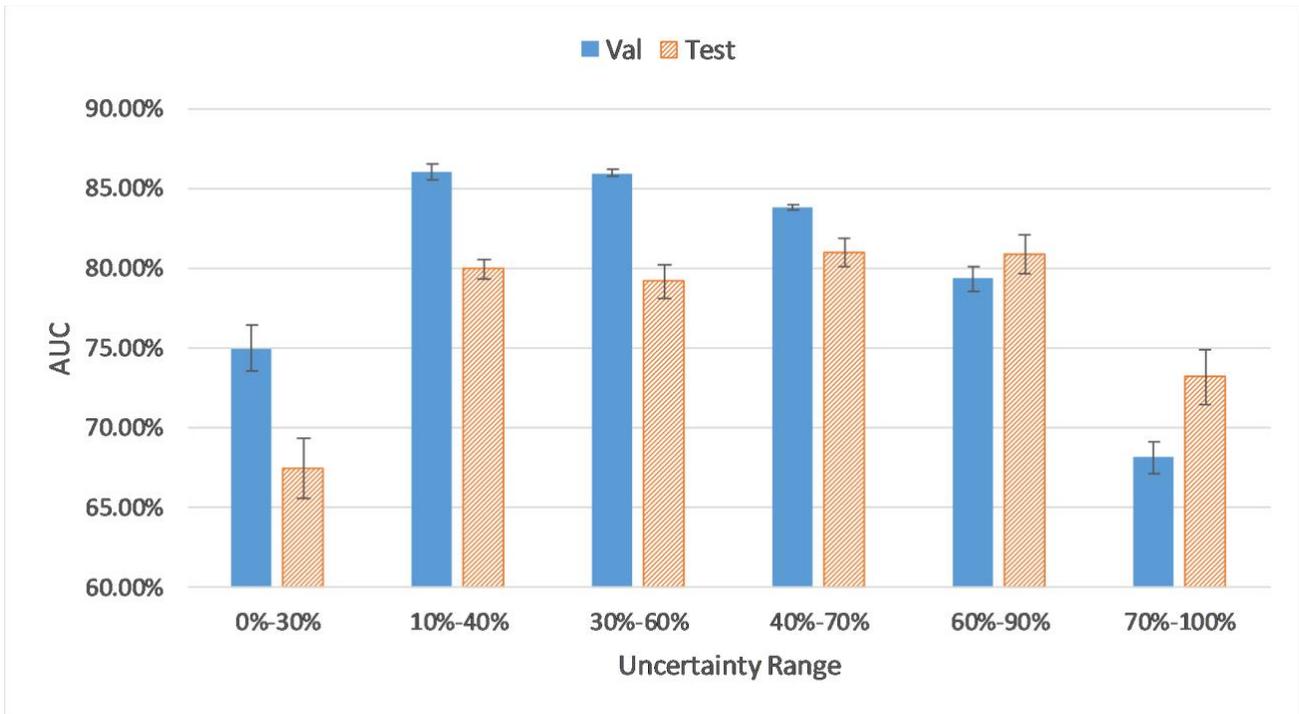

Figure 3. CNN performance on samples from different uncertainty ranges

The distribution and proportion of the hardest 10% samples and the easiest 10% samples in the entire uncertainty distribution are visualized in Figure 4. We hypothesize the top 10% uncertain examples are outliers, and the bottom 10% do not provide training value for the model, which will result in a poor model generalization.

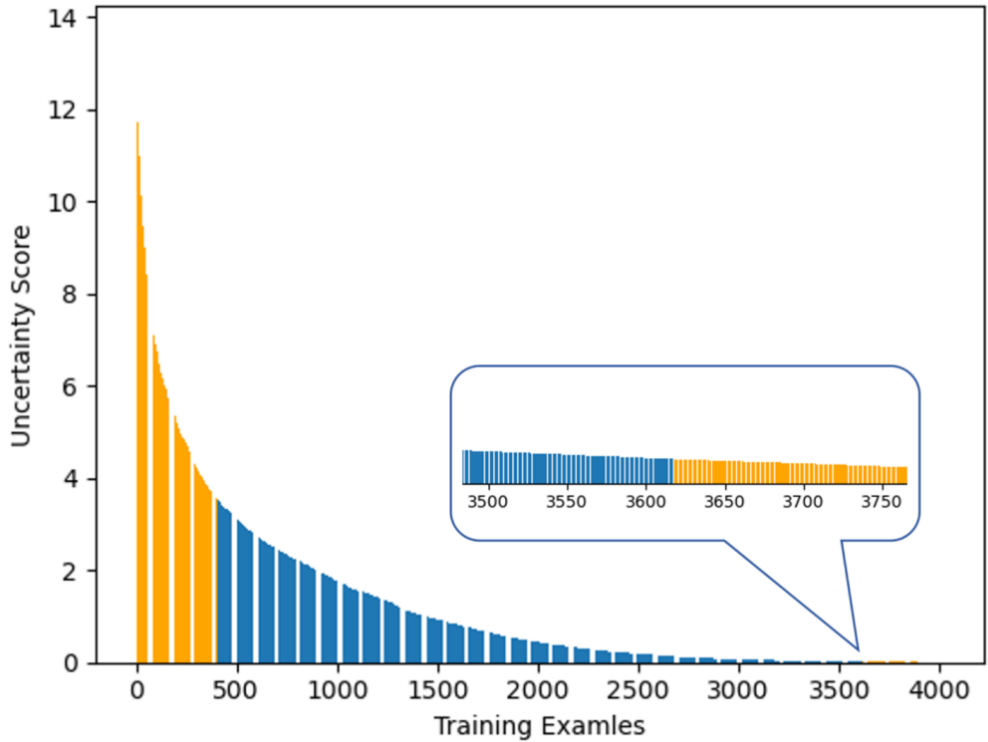

Figure 4. The distribution of 10% examples with the highest and lowest uncertainty scores

### 3.3 AUC Results of Uncertainty Sampling Method

In our uncertainty sampling algorithm, in order to train a model with better generalization, we discarded the top 10% and the bottom 10% training examples to eliminate outliers and least informative samples. Next, we randomly sampled 30% of the entire training cohort from the remaining dataset. We hypothesized that because this sample set did not include the top and bottom most uncertain and certain cases, it was the best informative and representative part of the dataset and hence, we used it to finetune a pretrained AlexNet, in order to achieve competitive model performance compared with using the whole training dataset.

Table 3 lists model classification performance based on the proposed uncertainty sampling method and compares it with the baseline in which we finetuned the pretrained AlexNet on the entire training dataset. Figure 5 illustrates contents of the Table 3.

It can be seen that our proposed uncertainty sampling method achieved similar classification performance on the validation dataset, and the AUC on the test set was 2.92% higher than the baseline AUC. Overall, the proposed method could save 40%-70% of labeling cost while maintaining high classification performance of the model.

Table 3. AUC results of the Proposed Method and Baseline AUC

| AUC (95%CI) | The Proposed Method | The Baseline |
|---|---|---|
| Validation dataset | 86.86% (86.48, 87.24) | 87.46% (87.11, 87.81) |
| Test dataset | 82.89% (81.87, 83.91) | 79.91% (78.95, 80.87) |

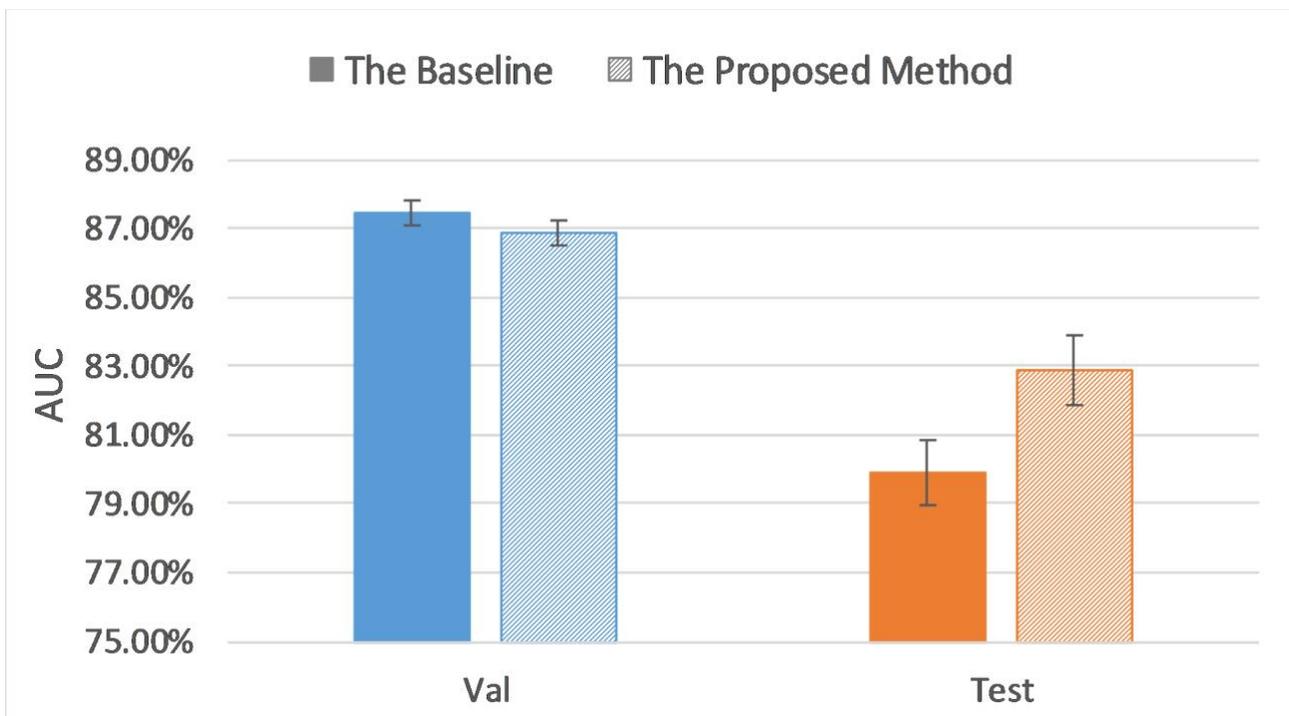

Figure 5. Comparison of AUC Results of the Proposed Method and Baseline

### 3.4 AUC Results of Uncertainty Sampling Method on Balanced Dataset

For the purpose of verifying the robustness of our proposed method, we further created a balanced dataset and applied uncertainty sampling method. In order to better control the variables, we did not change the way the training, validation, and test sets were divided. Rather, we changed the number of slices extracted from each patient's MRI scan. Because the ratio of the number of HGG patients (259 patients) and LGG patients (76 patients) is close to 3:1, the ratio of the number of HGG and LGG slices can be changed to 1:3 to form a balanced data set. Therefore, 30 slices were extracted from MRI scan instead of 20 slices for each LGG patient, and the number of MRI slices for every HGG patient reduced from 20 slices to 10 slices. This yielded a dataset of 4,870 2D 3-channel slices.

The baseline AUC was computed when the pretrained AlexNet was finetuned on the entire balanced training set, and the uncertainty sampling method was the same as described previously. As Table 4 and Figure 6 indicate, even on a balanced dataset, our proposed method achieved better classification performance than the baseline test AUC with significantly less annotations, which demonstrates robustness of our uncertainty sampling method.

Table 4. AUC Results of the Proposed Method and Baseline AUC on Balanced Dataset

| AUC (95%CI) | The Proposed Method | The Baseline |
|---|---|---|
| Validation dataset | 85.20% (84.88, 85.52) | 87.17% (86.87, 87.47) |
| Test dataset | 82.00% (81.18, 82.82) | 78.48% (77.60, 79.36) |

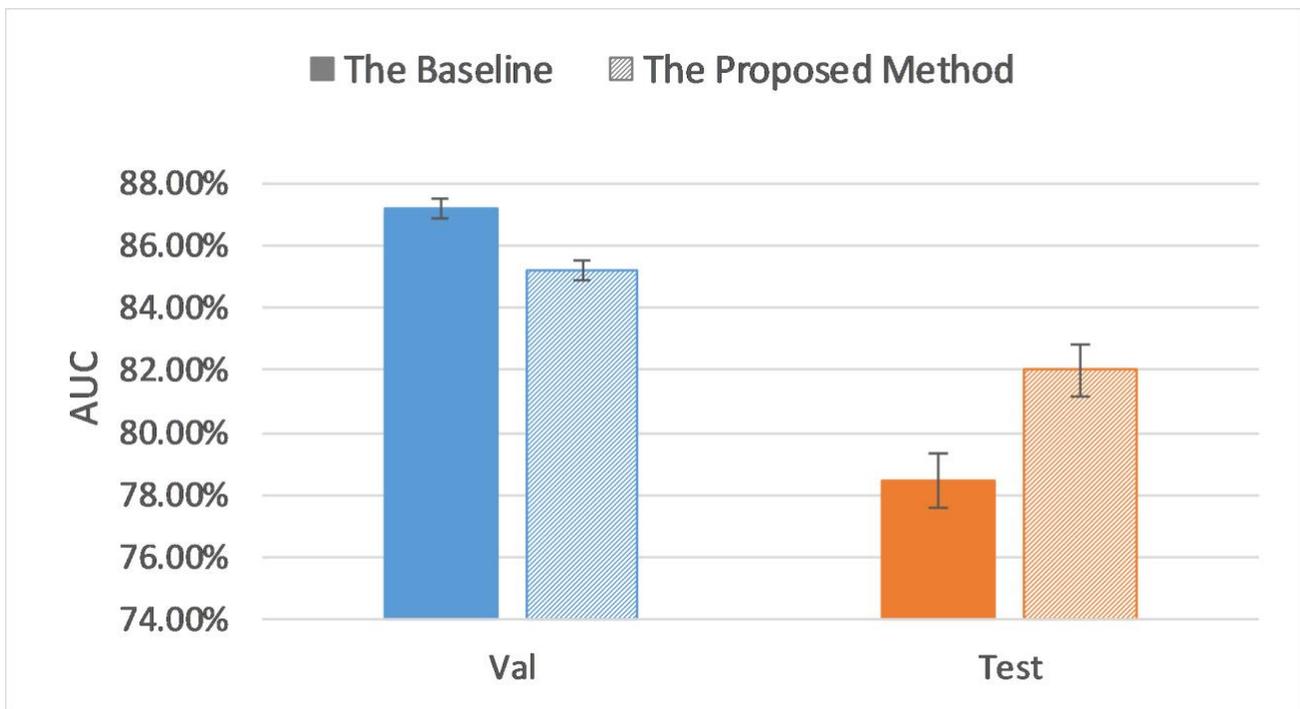

Figure 6. Comparison of AUC Results of the Proposed Method and Baseline on Balanced Dataset

### 3.5 Comparison of AUC Results of Uncertainty Sampling method and Random Sampling Method

In the previous sections, our sample size was fixed at 30% of the training dataset excluding the top and bottom most certain and uncertain samples. In this section we investigate effect of the sample size. In order to compare the efficacy of uncertainty sampling method and random sampling method, we finetuned the pretrained AlexNet on the fixed number of examples which were created using these two sampling methods.

In our random sampling method, we started with random sampling of 10% of the training cohort (N samples), then increased the number of samples by 10% of the training dataset (N) until it accounts for 80% of the total training set samples (8xN) (top and bottom 10% already removed). Thus, 8 sampled datasets with a sample size of 10%-80% (N to 8xN) of the total training set were obtained, with interval of 10% (N). For the uncertainty sampling method, we removed the top 10% and bottom 10% samples according to the sorted uncertainty list, and randomly selected a subset whose sample size is 10% of the total training cohort (N) from the remaining part of the dataset. Similar to the previous sampling process, we created 8 different datasets and conducted our experiments on them. Table 5 describes the details of correspondence between the proportion of sample size and the number of examples on imbalanced and balanced dataset.

Table 5. Correspondence between the Proportion of sample size and the number of examples on Imbalanced Dataset and Balanced Dataset

| Proportion of Sample Size | | 10% | 20% | 30% | 40% | 50% | 60% | 70% | 80% |
|---|---|---|---|---|---|---|---|---|---|
| Number of Examples | Imbalanced Dataset | 406 | 812 | 1218 | 1624 | 2030 | 2436 | 2842 | 3248 |
| | Balanced Dataset | 487 | 974 | 1461 | 1948 | 2435 | 2922 | 3409 | 3896 |

Figure 7 and Figure 8 show the visualizations of test AUC Results using the uncertainty sampling method and random sampling method on the imbalanced as well as the balanced datasets. In each figure, the solid and dash dotted lines indicate the AUC values obtained on the samples corresponding to the parameters of the horizontal axis, and the dotted lines represent the baseline AUC which were computed when the pretrained CNN was trained on the entire labeled training datasets. The two colors orange and blue in each figure represents AUC results calculated by the uncertainty sampling method and random sampling method, respectively.

As shown in Figure 7 and Figure 8, for both imbalanced and balanced datasets, our proposed method performs better than the random sampling method, and the AUC results are higher than the baseline on every proportion of sample size, which demonstrates the stability and robustness of our proposed uncertainty sampling strategy.

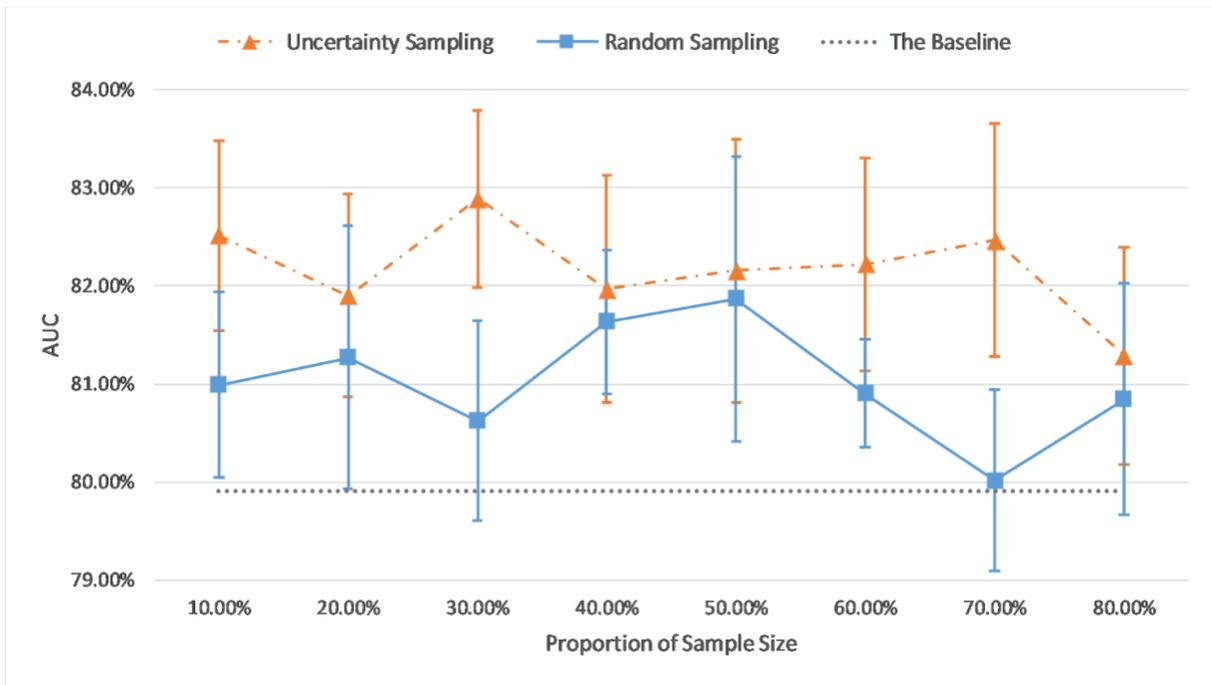

Figure 7. Comparison of Test AUC Results of the Uncertainty Sampling Method and Random Sampling Method on the Imbalanced Dataset

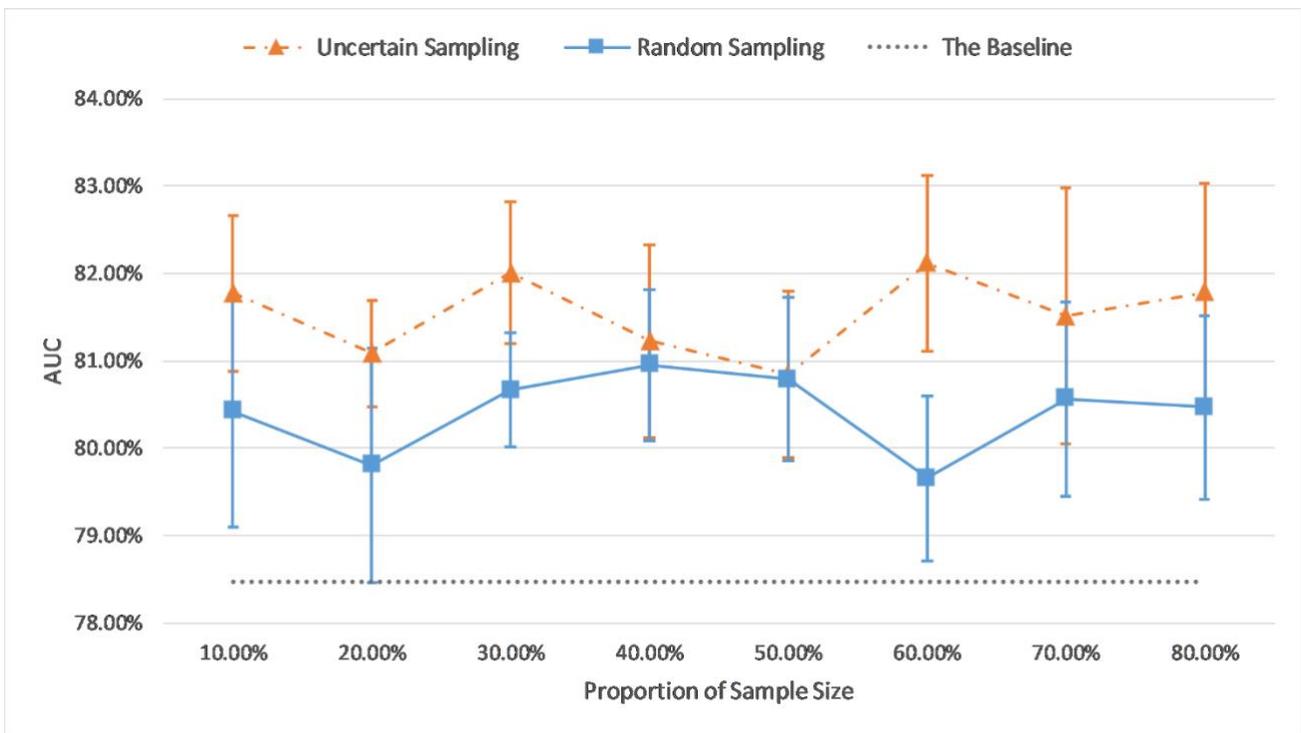

Figure 8. Comparison of Test AUC Results of the Uncertainty Sampling Method and Random Sampling Method on the Balanced Dataset

## 4  Discussion

Deep learning algorithms for detection of tumors in medical images require large annotated datasets for training. The annotation is usually done manually by subspecialty radiologists. The associated cost (time and expertise) is prohibitively high, which hinders success of AI in Medical Imaging. Transfer learning is a widely used approach which can transfer the knowledge that the model has learned on large datasets to the new recognition and classification tasks. Active learning algorithms have been investigated to train a competitive classifier with minimal annotation cost. In this work, we combine transfer learning and active learning to propose a novel uncertainty sampling method which can reduce the amount of required training samples while maintaining stability and robustness of CNN performance for brain tumor classification.

Our proposed sampling method selects samples with representativeness and informativeness by discarding subsets of training samples with the highest and lowest uncertainty scores. We set the proportion of discarded samples as 10%, because the top 10% examples with highest uncertainty and the bottom 10% samples with the lowest uncertainty resulted in a poor model generalization as shown in Figure 3. We then had multiple options for using the remaining 80% of the training dataset. Our experiments revealed that a sample as big as 30% of the dataset is the optimum choice (Figure 7 and Figure 8). With using 30% of the training dataset conditioned on excluding top and bottom 10% of our uncertainty list, uncertainty sampling method achieved AUC of 82.89% and 82.00% on the imbalanced and balanced datasets, respectively, which was comparable or better than the baseline AUC. Although for the balanced dataset the best sampling size would be 60%, given the slight difference between AUC results at 30% and 60% (82.00% vs 82.11%), we chose 30% to save a considerable amount of labeling costs and to be consistent with the imbalanced scenario. The proposed method can save 40%-70% of the labeling cost. We also compared our uncertainty method with random sampling and demonstrated that our proposed method outperforms random sampling. It should be noted that random sampling is inherently unstable compared to the proposed systematic sampling approach, and the results for random sampling are not reliable as they may not be repeatable.

In future works, the proposed method will be applied to other imaging modalities and cancer sites including prostate MRI and lung CT. In addition, the proportion of discarded samples and the proportion of samples selected from the remaining datasets can be further explored based on statistical knowledge.

## 5  Conclusion

A transfer learning based active learning framework can significantly reduce the size of required labeled training data while maintaining high accuracy of the classification of tumors in brain MRI.

**Data Availability Statement**

The datasets analyzed for this study can be found in the BRATS 2019 dataset https://www.med.upenn.edu/cbica/brats2019/data.html.

**Ethics Statement**

The study is based on a publicly available dataset from the Multimodal Brain Tumor Segmentation Challenge 2019 (BraTS, https://www.med.upenn.edu/cbica/brats-2019/).

**Author Contributions**

RH, KN, and FK contributed to the design of the concept, and the design and implementation machine learning modules. All authors contributed to the writing and reviewing of the paper and read and approved the final manuscript.

**Funding**

This research received funding support in part by China Scholarship Council, and Chair in Medical Imaging and Artificial Intelligence, a joint Hospital-University Chair between the University of Toronto, The Hospital for Sick Children, and the SickKids Foundation.

**Conflict of Interest**

The authors declare that the research was conducted in the absence of any commercial or financial relationships that could be construed as a potential conflict of interest.